\documentclass[twocolumn,showpacs,aps,epsfig,nofootinbib]{revtex4}

\usepackage{graphicx}
\usepackage{epstopdf}
\usepackage{latexsym}
\usepackage{amssymb}

\usepackage[center]{subfigure}

\begin{document}

 \newcommand{\bq}{\begin{equation}}
 \newcommand{\eq}{\end{equation}}
 \newcommand{\bqn}{\begin{eqnarray}}
 \newcommand{\eqn}{\end{eqnarray}}
 \newcommand{\nb}{\nonumber}
 \newcommand{\lb}{\label}
\newcommand{\PRL}{Phys. Rev. Lett.}
\newcommand{\PL}{Phys. Lett.}
\newcommand{\PR}{Phys. Rev.}
\newcommand{\CQG}{Class. Quantum Grav.}

\title{Stability, ghost,  and strong coupling   in  nonrelativistic general covariant theory of gravity with   $\lambda \not=1$} 

\author{Yongqing Huang}
\email{yongqing_huang@baylor.edu}

\author{Anzhong Wang}
\email{anzhong_wang@baylor.edu}

\affiliation{GCAP-CASPER, Physics Department, Baylor
University, Waco, TX 76798-7316, USA}

\date{\today}

\begin{abstract}

In this paper, we investigate  three important issues:  stability, ghost and strong coupling, in  the  Horava-Melby-Thompson 
setup of  the Horava-Lifshitz  theory  with $\lambda \not= 1$, generalized recently by da Silva.  We first develop the general 
linear scalar perturbations of the Friedmann-Robertson-Walker (FRW) universe with arbitrary spatial curvature, and find that 
an immediate by-product of the setup is that, in all the inflationary models described by a scalar field, the FRW universe is 
necessarily  flat.  Applying them to the case of the Minkowski background, we find that it is stable, and, similar to the  case 
$\lambda = 1$, the spin-0 graviton is eliminated.  The vector  perturbations vanish identically in the Minkowski background. 
Thus, similar to general relativity,  a free gravitational field in this setup is completely described by a spin-2 massless graviton 
even with $\lambda \not= 1$. We also study the ghost problem in the FRW background, and find explicitly the ghost-free 
conditions.  To study the strong coupling problem, we consider two different  kinds of  spacetimes all with the presence of 
matter, one is cosmological and the one is static.  We find that the coupling becomes strong for a process with energy higher 
than $M_{pl} |c_{\psi}|^{5/2}$ in the flat FRW background, and $M_{pl}|c_{\psi}|^{3}$  in a static weak gravitational field, 
where $|c_{\psi}| \equiv |(1-\lambda)/(3 \lambda -1)|^{1/2}$.

\end{abstract}


\maketitle

\section{Introduction}
\renewcommand{\theequation}{1.\arabic{equation}} \setcounter{equation}{0}

Recently, Horava proposed a quantum gravity theory \cite{Horava}, motivated by the Lifshitz 
theory in solid state physics \cite{Lifshitz}. 
The Horava-Lifshitz (HL) theory  is based on the perspective 
that Lorentz symmetry should appear as an emergent symmetry at long distances, but can be fundamentally 
absent at high energies \cite{Pav}. Along  this vein of thinking,    Horava considered systems whose scaling at short 
distances exhibits a strong anisotropy between space and time, 
\bq
\lb{1.1}
{\bf x} \rightarrow \ell {\bf x}, \;\;\;  t \rightarrow \ell^{z} t,
\eq
where $z \ge 3$,   in order for the theory to be power-counting 
renormalizable in $(3+1)$-dimensional spacetimes \cite{Visser}. At low energies, high-order curvature corrections become
negligible, and the theory is expected to  flow to $z = 1$, whereby the Lorentz invariance is ``accidentally restored." 
Such an anisotropy between  time and space can be easily realized  when one writes the metric
in the Arnowitt-Deser-Misner  (ADM) form  \cite{ADM},
 \bqn
 \lb{1.2}
ds^{2} &=& - N^{2}c^{2}dt^{2} + g_{ij}\left(dx^{i} + N^{i}dt\right)
     \left(dx^{j} + N^{j}dt\right), \nb\\
     & & ~~~~~~~~~~~~~~~~~~~~~~~~~~~~~~  (i, \; j = 1, 2, 3).~~~
 \eqn
 Under the rescaling (\ref{1.1})  
 with   $z  = 3$, a condition we shall assume in this paper,  the  lapse function   $N$, the shift vector $N^{i}$,
  and the 3-metric  $g_{ij}$ scale as, 
 \bq
 \lb{1.3}
  N \rightarrow  N ,\;  N^{i}
\rightarrow {\ell}^{-2} N^{i},\; g_{ij} \rightarrow g_{ij}.
 \eq
The gauge symmetry of the theory is the  foliation-preserving 
diffeomorphism, 
\bq
\lb{1.4}
\tilde{t} = t - f(t),\; \;\; \tilde{x}^{i}  =  {x}^{i}  - \zeta^{i}(t, {\bf x}),
\eq
often denoted by Diff($M, \; {\cal{F}}$),  for which $N,\; N^{i}$ and $g_{ij}$   transform as
\bqn
\lb{1.5}
\delta{N} &=& \zeta^{k}\nabla_{k}N + \dot{N}f + N\dot{f},\nb\\
\delta{N}_{i} &=& N_{k}\nabla_{i}\zeta^{k} + \zeta^{k}\nabla_{k}N_{i}  + g_{ik}\dot{\zeta}^{k}
+ \dot{N}_{i}f + N_{i}\dot{f}, \nb\\
\delta{g}_{ij} &=& \nabla_{i}\zeta_{j} + \nabla_{j}\zeta_{i} + f\dot{g}_{ij}, 
\eqn
where $\dot{f} \equiv df/dt,\;  \nabla_{i}$ denotes the covariant 
derivative with respect to the 3-metric $g_{ij}$,  $N_{i} = g_{ik}N^{k}$, and $\delta{g}_{ij} 
\equiv \tilde{g}_{ij}\left(t, x^k\right) - {g}_{ij}\left(t, x^k\right)$,
 etc. From these expressions one can see that   $N$ and   $N^{i}$ play the role of gauge fields of the Diff($M, \; {\cal{F}}$)
symmetry. Therefore, it is natural to assume that $N$ and $N^{i}$ inherit the same dependence on 
spacetime as the corresponding generators,  that is,
\bq
\lb{1.6}
N = N(t), \;\;\; N^{i} = N^{i}(t, x).
\eq
It  is   clear that this is preserved by Diff($M, \; {\cal{F}}$), and 
usually referred to as the {\em projectability condition} (Note that   the dynamical variables $g_{ij}$
 in general depend on both time and space, $g_{ij} = g_{ij}(t, x)$.).

Due to the restricted diffeomorphisms (\ref{1.4}), one more degree of freedom appears
in the gravitational sector - a spin-0 graviton. This is potentially dangerous, and needs to decouple  
in the IR, in order to be consistent with observations.   Whether this is possible or not is still an
open question \cite{Mukc}. In particular,  it was shown that   the spin-0 mode is not stable 
 in the original version of the HL theory  \cite{Horava} as well as in the Sotiriou, Visser 
and Weinfurtner (SVW) generalization \cite{SVW,WM,WWM}. But,   these
instabilities were all found in the Minkowski background. In the de Sitter
spacetime, it was shown that it is stable  
\cite{HWW}. So, one may take the latter 
as its legitimate  background  \cite{MGs}. However,
the strong coupling problem still exists \cite{SC,WWb}, although it might be   circumvented 
by the Vainshtein mechanism \cite{Vain}, as recently showed in the spherical  static  \cite{Mukc} and
 cosmological \cite{WWb} spacetimes.

To cure the above problems, various versions of the theory were proposed recently \cite{Sotiriou,BPS}. In particular,
Horava and Melby-Thompson (HMT) showed that one can eliminate the spin-0 graviton by introducing two auxiliary fields,
the $U(1)$ gauge field $A$ and the Newtonian pre-potentail $\varphi$, by   extending 
the  Diff($M, \; {\cal{F}}$) symmetry to include  a local $U(1)$ symmetry \cite{HMT}, 
\bq
\lb{symmetry}
 U(1) \ltimes {\mbox{Diff}}(M, \; {\cal{F}}).
 \eq
Under this extended symmetry,   the special status of time  maintains,  so that the anisotropic scaling (\ref{1.1})
with $z > 1$  is still  realized, whereby the UV behavior of the theory can be considerably improved.   
Under the local $U(1)$ symmetry,  the gravitational and gauge fields 
 transform as
\bqn
\lb{2.3}
\delta_{\alpha}A &=&\dot{\alpha} - N^{i}\nabla_{i}\alpha,\;\;\;
\delta_{\alpha}\varphi = - \alpha,\nb\\ 
\delta_{\alpha}N_{i} &=& N\nabla_{i}\alpha,\;\;\;
\delta_{\alpha}g_{ij} = 0 = \delta_{\alpha}{N},
\eqn
where $\alpha$ is   the generator  of the local $U(1)$ gauge symmetry. 
Under the Diff($M, \; {\cal{F}}$), $A$ and $\varphi$ transform as,
\bqn
\lb{2.2}
\delta{A} &=& \zeta^{i}\nabla_{i}A + \dot{f}A  + f\dot{A},\nb\\
\delta \varphi &=&  f \dot{\varphi} + \zeta^{i}\nabla_{i}\varphi.
\eqn
For details, we refer readers to \cite{HMT,WW}. 

As shown explicitly in \cite{Horava2}, the $U(1)$ symmetry pertains specifically to the case $\lambda = 1$,
where $\lambda$ is a coupling constant that 
characterizes the deviation of 
the kinetic part of action from the corresponding one given in general relativity (GR). It is exactly because 
of this deviation that causes all the   problems, including  ghost, instability and  strong coupling.
Therefore, it  was  considered as a remarkable feature of this nonrelativistic general covariant theory,  in which  
$\lambda$ is forced to be one. However, this claim  was soon challenged  by da Silva
\cite{Silva}, who  argued that the introduction of the   Newtonian pre-potential is so powerful that action with
  $\lambda \not=1$ also has the $U(1)$ symmetry \footnote{
It should be noted that even in the tree level we could have $\lambda = 1$, it is still subjected to quantum corrections. This is 
 in contrast to the relativistic case, where $\lambda = 1$ is protected by   the Lorentz 
symmetry,  
$\tilde{x}^{\mu} =   \tilde{x}^{\mu}(t, {\bf x}),\; (\mu = 0, 1, 2, 3)$,
even in the  quantum level. }.

Once the coupling constant $\lambda$ can be different from one,   the issues of instability, ghost and strong coupling plagued in other
versions of the HL theory all  rise again. 
In this paper,  we   investigate   these important  questions in detail in the framework of da Silva's generalization of the HMT setup. 
Specifically, in Sec. II we briefly review the theory by presenting  all the field equations and conservation laws 
when matter is  present. In Sec. III we study the Friedmann-Robertson-Walker (FRW) universe
 with any given spatial curvature, and derive the generalized 
Friedmann equation and conservation law of energy. An immediate by-product of the setup is that, in all the inflationary models described by a scalar field, 
the FRW universe is necessarily  flat.  In Sec. IV,  we develop the general formulas for the linear
scalar perturbations of the FRW universe. Applying these formulas to the Minkowski background
in Sec.V,  we study the stability problem, and show explicitly that it is stable and the spin-0
graviton is   eliminated even for $\lambda \not=1$. This conclusion is the same as that obtained by
 da Silva for the maximal symmetric spacetimes with detailed balance condition, in which the Minkowski spacetime is
not a solution of the theory \cite{Silva}. In Sec. VI, we study the ghost and strong coupling problems, and derive the ghost-free
conditions in terms of $\lambda$. 
To study the strong coupling problem, 
we consider two different  kinds of  spacetimes all filled with matter  \footnote{ Note that,  to  count  the number of 
the    degrees of the  propagating gravitational modes,  one needs to consider free gravitational fields. 
Another way to  count the degrees of the freedom is to  study the structure  of  the Hamiltonian constraints  \cite{HMT,Kluson}.
On the other hand, to study the ghost and strong coupling problems, one needs to consider  the cases in which the gravitational perturbations are different from zero. 
In this paper, this is realized by the presence of matter fields. That is, it is the matter that produces the gravitational perturbations. Clearly, this does not contradict 
to the conclusion that the spin-0 mode is eliminated in such a setup.  A similar situation also 
happens in GR, in which gravitational scalar perturbations of the FRW universe in general do not vanish, although the only degrees of the freedom
of the gravitational sector  are the spin-2 massless gravitons.}: 
(a) spacetimes in which  the flat 
FRW universe   can be considered as their zero-order approximations; and (b)  spherical statics spacetimes in 
which the Minkowski spacetime can be considered as their zero-order approximations.  We find  that the strong coupling
problem indeed exists in both kinds of spacetimes.    It should be noted that strong coupling itself
is not a problem, as long as the theory is consistent with observations. In fact, several well-known
 theories are strong coupling \cite{Pol}. Interestingly enough, the strong coupling in the 
Dvali-Gabadadze-Porrati  braneworld model  even helps to screen   the spin-0 mode so that the models are consistent with solar system tests \cite{Koyama}. 
Finally, in Sec. VII we present our main conclusions and discussing remarks. An appendix is also included, in which, among other things, 
 the kinetic part of the  action and coupling coefficients are given. 

Before proceeding further, we would like to note that  in \cite{WW} we studied the HMT
setup   where $\lambda = 1$. In addition, static spacetimes  were also studied recently \cite{AP,GSW}, while its Hamiltonian structure and some possible generalizations
were  investigated  in \cite{Kluson}.  In all of these investigations  $\lambda = 1$ was assumed.  Thus, in this paper we are mainly concerned with  $\lambda \not=1$. 

Moreover, cosmology and black hole physics in other versions of the HL theory have been  intensively studied recently,  and it becomes very difficult to review 
all those important works here.   Instead, we simply refer readers to {\cite{HWW,Sotiriou,Mukc,Padilla} for detail.

\section{Nonrelativisitc general covariant HL theory with any $\lambda$}

\renewcommand{\theequation}{2.\arabic{equation}} \setcounter{equation}{0}

For any given coupling constant   $\lambda$,   the total action can be written as \cite{HMT,WW,Silva},
 \bqn \lb{2.4}
S &=& \zeta^2\int dt d^{3}x N \sqrt{g} \Big({\cal{L}}_{K} -
{\cal{L}}_{{V}} +  {\cal{L}}_{{\varphi}} +  {\cal{L}}_{{A}} +  {\cal{L}}_{{\lambda}} \nb\\
& & ~~~~~~~~~~~~~~~~~~~~~~ \left. + {\zeta^{-2}} {\cal{L}}_{M} \right),
 \eqn
where $g={\rm det}\,g_{ij}$, and
 \bqn \lb{2.5}
{\cal{L}}_{K} &=& K_{ij}K^{ij} -   \lambda K^{2},\nb\\
{\cal{L}}_{\varphi} &=&\varphi {\cal{G}}^{ij} \Big(2K_{ij} + \nabla_{i}\nabla_{j}\varphi\Big),\nb\\
{\cal{L}}_{A} &=&\frac{A}{N}\Big(2\Lambda_{g} - R\Big),\nb\\
{\cal{L}}_{\lambda} &=& \big(1-\lambda\big)\Big[\big(\nabla^{2}\varphi\big)^{2} + 2 K \nabla^{2}\varphi\Big].
 \eqn
Here 
$\Lambda_{g}$ is a    coupling constant, and the
Ricci and Riemann terms all refer to the three-metric $g_{ij}$, and 
 \bqn \lb{2.6}
K_{ij} &=& \frac{1}{2N}\left(- \dot{g}_{ij} + \nabla_{i}N_{j} +
\nabla_{j}N_{i}\right),\nb\\
{\cal{G}}_{ij} &=& R_{ij} - \frac{1}{2}g_{ij}R + \Lambda_{g} g_{ij}.
 \eqn
${\cal{L}}_{M}$ is the
matter Lagrangian density, which in general is a function of all the dynamical variables,
$U(1)$ gauge field, and the Newtonian prepotential, i.e., 
$
{\cal{L}}_{M} = {\cal{L}}_{M}\big(N, \; N_{i}, \; g_{ij}, \; \varphi,\; A; \; \chi\big)$,
where $\chi$ denotes collectively the matter fields. ${\cal{L}}_{{V}}$ is an arbitrary Diff($\Sigma$)-invariant local scalar functional
built out of the spatial metric, its Riemann tensor and spatial covariant derivatives, without the use of time derivatives.

Note the difference between the notations used here and the ones used in \cite{HMT,Silva} \footnote{In particular, we
have $K_{ij} = - K_{ij}^{HMT},\; \Lambda_{g} = \Omega^{HMT},\; \varphi = - \nu^{HMT}, {\cal{G}}_{ij} = \Theta_{ij}^{HMT}$, 
where quantities with the super-indice ``HMT" are those used in \cite{HMT,Silva}.}. In this paper, without further explanations,
we shall use directly  the notations and conventions defined in \cite{WM} and \cite{WW}, which will be referred, respectively, 
to as Paper I and Paper II.   However, in order to have the present paper as independent as possible, it is difficult to avoid 
repeating the same materials, although we shall try to limit it to its minimum. 

In \cite{SVW}, by assuming that the highest order derivatives are six, the minimum in order to have the theory  to be power-counting 
renormalizable \cite{Visser}, and that  the theory  preserves 
the parity, SVW constructed the most general form of  ${\cal{L}}_{{V}}$, 
 \bqn \lb{2.5a} 
{\cal{L}}_{{V}} &=& \zeta^{2}g_{0}  + g_{1} R + \frac{1}{\zeta^{2}}
\left(g_{2}R^{2} +  g_{3}  R_{ij}R^{ij}\right)\nb\\
& & + \frac{1}{\zeta^{4}} \left(g_{4}R^{3} +  g_{5}  R\;
R_{ij}R^{ij}
+   g_{6}  R^{i}_{j} R^{j}_{k} R^{k}_{i} \right)\nb\\
& & + \frac{1}{\zeta^{4}} \Big[g_{7}R\nabla^{2}R +  g_{8}
\left(\nabla_{i}R_{jk}\right)
\left(\nabla^{i}R^{jk}\right)\Big],  ~~~~
 \eqn 
 where the coupling  constants $ g_{s}\, (s=0, 1, 2,\dots 8)$  are all dimensionless. The relativistic limit in the IR
 requires $g_{1} = -1$ and $\zeta^2 = 1/(16\pi G)$ \cite{SVW}. 

Then, it can be shown that the  Hamiltonian and momentum constraints are given respectively by,
 \bqn 
 \lb{eq1}
& & \int{ d^{3}x\sqrt{g}\left[{\cal{L}}_{K} + {\cal{L}}_{{V}} - \varphi {\cal{G}}^{ij}\nabla_{i}\nabla_{j}\varphi 
- \big(1-\lambda\big)\big(\nabla^{2}\varphi\big)^{2}\right]}\nb\\
& & ~~~~~~~~~~~~~~~~~~~~~~~~~~~~~
= 8\pi G \int d^{3}x {\sqrt{g}\, J^{t}},\\
\lb{eq2}
& & \nabla^{j}\Big[\pi_{ij} - \varphi  {\cal{G}}_{ij} - \big(1-\lambda\big)g_{ij}\nabla^{2}\varphi \Big] = 8\pi G J_{i},
 \eqn
where 
 \bqn 
  \lb{eq2b}
  J^{t} &\equiv& 2 \frac{\delta\left(N{\cal{L}}_{M}\right)}{\delta N},\nb\\
   \pi_{ij} &\equiv& 
   - K_{ij} +  \lambda K g_{ij},\nb\\
 J_{i} &\equiv& - N\frac{\delta{\cal{L}}_{M}}{\delta N^{i}}.
 \eqn
 
Variation of the action (\ref{2.4}) with respect to   $\varphi$ and $A$ yield, respectively, 
\bqn
\lb{eq4a}
& & {\cal{G}}^{ij} \Big(K_{ij} + \nabla_{i}\nabla_{j}\varphi\Big) + \big(1-\lambda\big)\nabla^{2}\Big(K + \nabla^{2}\varphi\Big) \nb\\
& & ~~~~~~~~~~~~~~~~~~~~~~~~~~~~~~~~ = 8\pi G J_{\varphi}, \\
\lb{eq4b}
& & R - 2\Lambda_{g} =   8\pi G J_{A},
\eqn
where
\bq
\lb{eq5}
J_{\varphi} \equiv - \frac{\delta{\cal{L}}_{M}}{\delta\varphi},\;\;\;
J_{A} \equiv 2 \frac{\delta\left(N{\cal{L}}_{M}\right)}{\delta{A}}.
\eq
On the other hand,  the dynamical equations now read,
 \bqn \lb{eq3}
&&
\frac{1}{N\sqrt{g}}\Bigg\{\sqrt{g}\Big[\pi^{ij} - \varphi {\cal{G}}^{ij} - \big(1-\lambda\big) g^{ij} \nabla^{2}\varphi\Big]\Bigg\}_{,t} 
\nb\\
& &~~~ = -2\left(K^{2}\right)^{ij}+2\lambda K K^{ij} \nb\\
& &  ~~~~~ + \frac{1}{N}\nabla_{k}\left[N^k \pi^{ij}-2\pi^{k(i}N^{j)}\right]\nb\\
& &  ~~~~~ - 2\big(1-\lambda\big) \Big[\big(K + \nabla^{2}\varphi\big)\nabla^{i}\nabla^{j}\varphi + K^{ij}\nabla^{2}\varphi\Big]\nb\\
& & ~~~~~ + \big(1-\lambda\big) \Big[2\nabla^{(i}F^{j)}_{\varphi} - g^{ij}\nabla_{k}F^{k}_{\varphi}\Big]\nb\\
& & ~~~~~ +  \frac{1}{2} \Big({\cal{L}}_{K} + {\cal{L}}_{\varphi} + {\cal{L}}_{A} + {\cal{L}}_{\lambda}\Big) g^{ij} \nb\\
& &  ~~~~~    + F^{ij} + F_{\varphi}^{ij} +  F_{A}^{ij} + 8\pi G \tau^{ij},
 \eqn
where $\left(K^{2}\right)^{ij} \equiv K^{il}K_{l}^{j},\; f_{(ij)}
\equiv \left(f_{ij} + f_{ji}\right)/2$, and
 \bqn
\lb{eq3a} 
F^{ij} &\equiv&
\frac{1}{\sqrt{g}}\frac{\delta\left(-\sqrt{g}
{\cal{L}}_{V}\right)}{\delta{g}_{ij}}
 = \sum^{8}_{s=0}{g_{s} \zeta^{n_{s}}
 \left(F_{s}\right)^{ij} },\nb\\
F_{\varphi}^{ij} &=&  \sum^{3}_{n=1}{F_{(\varphi, n)}^{ij}},\nb\\
F_{\varphi}^{i} &=&  \Big(K + \nabla^{2}\varphi\Big)\nabla^{i}\varphi + \frac{N^{i}}{N} \nabla^{2}\varphi, \nb\\
F_{A}^{ij} &=& \frac{1}{N}\left[AR^{ij} - \Big(\nabla^{i}\nabla^{j} - g^{ij}\nabla^{2}\Big)A\right],\nb\\ 
 \eqn
with 
$n_{s} =(2, 0, -2, -2, -4, -4, -4, -4,-4)$. The
stress 3-tensor $\tau^{ij}$ is defined as
 \bq \label{tau}
\tau^{ij} = {2\over \sqrt{g}}{\delta \left(\sqrt{g}
 {\cal{L}}_{M}\right)\over \delta{g}_{ij}},
 \eq
and the geometric 3-tensors $ \left(F_{s}\right)_{ij}$ and $F_{(\varphi, n)}^{ij}$ are  given in Paper II.

The matter components $(J^{t}, \; J^{i},\; J_{\varphi},\; J_{A},\; \tau^{ij})$ satisfy the
conservation laws,
 \bqn \lb{eq5a} & &
 \int d^{3}x \sqrt{g} { \left[ \dot{g}_{kl}\tau^{kl} -
 \frac{1}{\sqrt{g}}\left(\sqrt{g}J^{t}\right)_{, t}  
 +   \frac{2N_{k}}  {N\sqrt{g}}\left(\sqrt{g}J^{k}\right)_{,t}
  \right.  }   \nb\\
 & &  ~~~~~~~~~~~~~~ \left.   - 2\dot{\varphi}J_{\varphi} -  \frac{A} {N\sqrt{g}}\left(\sqrt{g}J_{A}\right)_{,t}
 \right] = 0,\\
\lb{eq5b} & & \nabla^{k}\tau_{ik} -
\frac{1}{N\sqrt{g}}\left(\sqrt{g}J_{i}\right)_{,t}  - \frac{J^{k}}{N}\left(\nabla_{k}N_{i}
- \nabla_{i}N_{k}\right)   \nb\\
& & \;\;\;\;\;\;\;\;\;\;\;- \frac{N_{i}}{N}\nabla_{k}J^{k} + J_{\varphi} \nabla_{i}\varphi - \frac{J_{A}}{2N} \nabla_{i}A
 = 0.
\eqn

\section{Cosmological Models}

\renewcommand{\theequation}{3.\arabic{equation}} \setcounter{equation}{0}

The homogeneous and isotropic universe is described by,
\bq
\lb{3.1}
\bar{N} = 1,\;\; \bar{N}_{i} = 0,\;\; 
\bar{g}_{ij} = a^{2}(t)\gamma_{ij},
\eq
where 
$\gamma_{ij}={\delta_{ij}}{\left(1 + \frac{1}{4}kr^{2}\right)^{-2}}$,
with $r^{2} \equiv x^2 + y^2 + z^2,\; k = 0, \pm 1$. As in  Paper I, 
we use symbols with bars to denote the quantities of background. 
Using the $U(1)$ gauge freedom of Eq.(\ref{2.3}), on the other hand, we can always set
\bq
\lb{3.3}
 \bar{\varphi} = 0.
\eq
Then, we find 
\bqn
\lb{3.4}
\bar K_{ij} &=& - a^{2}H \gamma_{ij}, \;\;\; \bar R_{ij} = 2k\gamma_{ij}, \nb\\
\bar F^{ij}_{A} &=& \frac{2k\bar A}{a^{4}} \gamma^{ij}, \;\;\; \bar F^{ij}_{\varphi} = 0,\;\;\; \bar F^{i}_{\varphi} = 0,\nb\\
\bar F^{ij} &=& \frac{\gamma^{ij}}{a^{2}}\left( - \Lambda + \frac{k}{a^{2}} 
+ \frac{2\beta_1k^{2}}{a^{4}}  + \frac{12\beta_2k^{3}}{a^{6}}\right),
\eqn
where $H = \dot{a}/a, \;  \Lambda \equiv \zeta^{2} g_{0}/2$, and  
\bq
\lb{3.5}
\beta_1 \equiv \frac{3g_{2} + g_{3}} {\zeta^{2}},\;\;\; 
\beta_2 \equiv \frac{9g_{4} + 3g_{5} + g_{6}}{\zeta^{4}}.
\eq
Hence,  we obtain 
 \bqn
\lb{3.6}
\bar{\cal{L}}_{K} &=&  3\big(1-3\lambda\big) H^{2},\;\; \bar{\cal{L}}_{\varphi} = 0 = \bar{\cal{L}}_{\lambda}, \nb\\
\bar{\cal{L}}_{A} &=& 2\bar A\Big(\Lambda_{g} - \frac{3k}{a^{2}}\Big), \nb\\
\bar{\cal{L}}_{V} &=& 2\Lambda - \frac{6k}{a^{2}} +
\frac{12\beta_1k^{2}}{a^{4}}  + \frac{24\beta_2k^{3}}{a^{6}}.
 \eqn
It can be shown that the super-momentum constraint (\ref{eq2}) is satisfied identically for
$\bar{J}^i = 0$, while the Hamiltonian constraint 
(\ref{eq1}) yields,
 \bq \lb{3.7a}
\frac{1}{2}\big(3\lambda - 1\big)H^{2} + \frac{k}{a^{2}} =
\frac{8\pi G}{3} \bar\rho+ \frac{\Lambda}{3} 
+ \frac{2\beta_1k^{2}}{a^{4}} + \frac{4\beta_2k^{3}}{a^{6}},
 \eq
where $\bar{J}^t \equiv -2\bar\rho$.
 On the other hand, Eqs.(\ref{eq4a}) and (\ref{eq4b}) give, respectively,
 \bqn
 \lb{3.8a}
 & & H\left(\Lambda_{g} - \frac{k}{a^{2}}\right) = - \frac{8\pi G}{3} \bar J_{\varphi},\\
 \lb{3.8b}
 & & \frac{3k}{a^{2}} -  \Lambda_{g}=  4\pi G \bar J_{A},
 \eqn
while the dynamical equation (\ref{eq3}) reduces to 
 \bqn
 \lb{3.7b}
 \frac{1}{2}\big(3\lambda - 1\big)\frac{\ddot{a}}{a} &=&  - {4\pi G\over
3}(\bar\rho+3 \bar p)+ {1\over3} \Lambda - \frac{2\beta_{1}k^{2}}{a^{4}}\nb\\
& &
  - \frac{8\beta_{2}k^{3}}{a^{6}}
+ \frac{1}{2}\bar A\left(\frac{k}{a^{2}} - \Lambda_{g}\right),
 \eqn
where   $\bar\tau_{ij} =  \bar p\,
\bar g_{ij}$.

The conservation law of momentum (\ref{eq5b}) is satisfied identically, while 
the one of energy (\ref{eq5a}) reduces to, 
 \bq \lb{3.8}
\dot{{\bar\rho}} + 3H \left(\bar\rho + \bar p \right) = \bar A \bar J_{\varphi}.
 \eq
It is interesting to note that the energy of matter is not conserved in general, due to its interaction with
the gauge field $\bar A$ and the Newtonian pre-potential $\bar\varphi$. This might have profound 
implications in cosmology.

\section{Cosmological Perturbations }

\renewcommand{\theequation}{4.\arabic{equation}} \setcounter{equation}{0}

As in Papers I and II, when we consider perturbations, we turn to the conformal time $\eta$, where $\eta = \int{dt/a(t)}$.
Under this coordinate transformation, the gravitational and gauge fields transfer as, 
\bqn
\lb{4.00}
 N &=& a\tilde{N}, \;\;\; N^{i} = a \tilde{N}^{i},\; \;\; g_{ij} = \tilde{g}_{ij},\nb\\
 A &=& a \tilde{A},  \;\;\; \varphi = \tilde{\varphi}, 
 \eqn
 where the quantities with tildes are the ones defined in the coordinates ($t, x^{i}$). With these in mind, 
we write the linear scalar perturbations of the metric   in terms of the conformal time $\eta$ as,
\bqn
\lb{4.0a}
\delta{N} &=& a \phi,\;\;\; \delta{N}_{i} = a^{2}B_{|i},\nb\\
\delta{g}_{ij} &=& -2a^{2}\big(\psi \gamma_{ij} - E_{|ij}\big),\nb\\
A &=& \hat{A} + \delta{A},\;\;\; \varphi = \hat{\varphi} + \delta\varphi,
\eqn
where
\bq
\lb{4.01}
\hat{A} = a \bar{A},\;\;\; \hat{\varphi} = \bar{\varphi},
\eq
and $\bar{A}$ and $\bar{\varphi}$ are the gauge fields of the background, given in the last section in the ($t, x^{i}$) coordinates. 
Under the gauge transformations (\ref{1.4}), we find that
\bqn
\lb{4.0b}
\tilde{\phi} &=& \phi - {\cal{H}}\xi^{0} - \xi^{0'},\;\;\;
\tilde{\psi} = \psi +  {\cal{H}}\xi^{0},\nb\\
\tilde{B} &=& B +  \xi^{0} - \xi',\;\;\;
\tilde{E} = E -   \xi,\nb\\
\tilde{\delta\varphi} &=& \delta\varphi - \xi^0 \hat{\varphi}',\;\;\;
\tilde{\delta{A}} = \delta{A} - \xi^0 \hat{A}' - \xi^{0'} \hat{A}, ~~~
\eqn
where $f = - \xi^0,\; \zeta^i = - \xi^{|i}, \;  {\cal{H}} \equiv a'/a$, and a prime denotes the ordinary derivative with respect 
to $\eta$. Under the $U(1)$ gauge transformations, on the other hand,
we find that
\bqn
\lb{4.0c}
\tilde{\phi} &=& \phi,\;\;\; 
\tilde{E} = E,\;\;\; 
\tilde{\psi} = \psi,\;\;\;
\tilde{B} =  B - \frac{\epsilon}{a}, \nb\\
\tilde{\delta\varphi} &=& \delta\varphi + \epsilon,\;\;\;
\tilde{\delta{A}} = \delta{A} - \epsilon', ~~~
\eqn
where $\epsilon = - \alpha$. Then, the  gauge transformations of the whole group $ U(1) \ltimes {\mbox{Diff}}(M, \; {\cal{F}})$
will be the linear combination of the above two. Since we have six unknown and three  arbitrary functions,  the total number of the
gauge-invariants of  $U(1) \ltimes {\mbox{Diff}}(M, \; {\cal{F}})$ is $N = 6 - 3 = 3$. These gauge-invariants can be constructed as,
\bqn
\lb{4.0d}
\Phi &=& \phi + \frac{a}{a - \hat{\varphi}'}\Bigg[\frac{{\delta\varphi}'}{a} + {\cal{H}}\big(B - E'\big)
+ \big(B - E'\big)'\Bigg]\nb\\
& & +  \frac{1}{\big(a - \hat{\varphi}'\big)^2}\big(\hat{\varphi}'' - {\cal{H}}\hat{\varphi}'\big)
\Big[\delta\varphi + a\big(B - E'\big)\Big],\nb\\
\Psi &=& \psi - \frac{{\cal{H}}}{a - \hat{\varphi}'}\Big[\delta\varphi +a\big(B - E'\big)\Big],\nb\\
\Gamma &=& \delta{A} + \Bigg[\frac{a + \hat{A}}{a - \hat{\varphi}'}\delta\varphi 
+ \frac{a(\hat{A} + \hat{\varphi}')}{a - \hat{\varphi}'} \big(B - E'\big)\Bigg]'.
\eqn

Using the $U(1)$ gauge freedom (\ref{4.0c}), we shall  set
\bq
\lb{4.0e}
{\delta\varphi} = 0.
\eq
This  choice   completely fixes the $U(1)$ gauge. 
Then, considering Eq.(\ref{3.3}), we find that the above expressions reduce to
\bqn
\lb{4.0f}
\Phi &=& \phi +  {\cal{H}}\big(B - E'\big)
+ \big(B - E'\big)', \nb\\
\Psi &=& \psi -  {\cal{H}} \big(B - E'\big),\nb\\
\Gamma &=& \delta{A} + \Big[\hat{A} \big(B - E'\big)\Big]',\;
(\hat\varphi = \delta\varphi = 0).
\eqn
The expressions for $\Phi$ and $\Psi$ now take precisely the same forms as those defined in
Paper I, which  are also identical to those given in GR  \cite{MW09}.
In Papers I and II, the  quasi-longitudinal gauge,
\bq
\lb{4.0g}
{\phi} = 0 = {E},
\eq
was imposed.   In  this paper, we shall adopt this gauge  for the metric perturbations, and   the gauge of
Eq.(\ref{4.0e})
for the Newtonian pre-potential. We shall   refer them as the ``generalized" quasi-longitudinal gauge, or 
simply the quasi-longitudinal gauge.

 Then, to first-order  the Hamiltonian and momentum constraints become,  respectively,
 \bqn \lb{4.4}
& & \int \sqrt{\gamma}d^{3}x\Bigg[\left(\vec\nabla^2+3k\right)\psi
- \frac{(3\lambda -1){\cal H}}{2}
\left(\vec\nabla^2 B + 3\psi'\right) \nb\\
& &~~~ -2 k\Big(\frac{2\beta_{1}}{a^{2}} +
\frac{6\beta_{2}k}{a^{4}}
+ \frac{3g_{7}}{\zeta^{4}a^{4}}\vec\nabla^2\Big)\left(\vec\nabla^2+3k\right)\psi\nb\\
& &   ~~~~~~~~~~~~~~~ -{4\pi G a^{2}}\delta{\mu}\Big]=0,\\
 \lb{4.5}
& & (3\lambda -1){\psi}' - 2kB - (1- \lambda)\vec\nabla^2B = 8\pi G a {q}, 
 \eqn
where $\delta\mu \equiv -\delta{J^{t}}/2$ and $\delta{J}^{i} \equiv a^{-2} q^{|i}$. 
On the other hand, the linearized equations (\ref{eq4a}) and (\ref{eq4b}) reduce, respectively,  to 
 \bqn
 \lb{4.6a}
& &  \Big(\Lambda_{g} - \frac{k}{a^{2}}\Big)\Big[\vec{\nabla}^{2}B + 3 \big(\psi' + 2{\cal{H}}\psi\big)\Big]\nb\\
 & &~~~~ + \frac{2{\cal{H}}}{a^{2}}\Big[\vec{\nabla}^{2}\psi + 3 \big(2k  - a^{2}\Lambda_{g}\big)\psi\Big] \nb\\
 & & ~~~~ + \frac{1-\lambda}{a^{2}}\vec{\nabla}^{2}\Big(\vec{\nabla}^{2}B + 3\psi' \Big) = 8\pi G a \delta J_{\varphi},~~~~~~\\
 \lb{4.6b}
& & \vec{\nabla}^{2}\psi + 3k\psi = 2\pi G a^{2} \delta J_{A}.
 \eqn

The linearly perturbed dynamical equations can be divided into the trace and traceless parts. The trace part  reads,
 \bqn
\lb{4.7a}
 \psi'' &+& 2{\cal{H}}\psi'  - {\cal{F}}\psi - \frac{1}{3(3\lambda-1)}
 \gamma^{ij}\delta{F}_{ij}\nb\\
&-& \frac{1}{3a(3\lambda -1)}\Big(2\vec\nabla^2 - 3k + 3\Lambda_{g}a^{2}\Big)\delta{A}\nb\\
&+& \frac{2\hat{A}}{3a(3\lambda -1)}\Big(\vec\nabla^2 +6k - 3\Lambda_{g}a^{2}\Big)\psi\nb\\
 &+& {1\over 3}\vec\nabla^2\left(B'+2 {\cal H}B \right)
 = { 8\pi Ga^{2}\over (3\lambda -1)} \delta{\cal P},
\eqn 
where 
 \bqn
 \lb{4.8}
 {\cal{F}} &=& \frac{2a^2}{3\lambda -1}\left(-\Lambda + {k \over a^2} +{2\beta_1k^2
 \over a^4} +{12\beta_2 k^3 \over a^6 }\right),\nb\\
 \delta{\tau}^{ij} &=& \frac{1}{a^{2}}\left[\left(\delta{\cal P} +
2\bar{p}\psi\right)\,\gamma^{ij} + {\Pi}^{|\langle ij\rangle}\right], \nb\\
f_{|\langle ij \rangle} &\equiv& f_{|ij}-{1\over 3}
\gamma_{ij}\vec{\nabla}^{2}f,
 \eqn
 and $\delta F_{ij}=\sum g_s \zeta^{n_s}\delta (F_s)_{ij}$, with
$\delta(F_s)_{ij}$ given by Eq.~(A1) in Paper I.  
 The traceless part   is given by
 \bqn
 \label{4.7b}
& & \Big(B' +2{\cal H}B\Big)_{|\langle ij \rangle} +\delta
F_{\langle ij \rangle} - \frac{1}{a}\left(\delta{A} - {\hat{A}}\psi\right)_{|\langle ij \rangle}\nb\\
& & ~~~~~~~~~~~~~~~~~~~~~~  =-8\pi G a^{2} \Pi_{|\langle ij \rangle}.
 \eqn
 
 To first order, the conservation laws 
 (\ref{eq5a}) and (\ref{eq5b}), on the other hand, take the forms,
 \bqn \lb{4.9a}
& & \int \sqrt{\gamma} d^{3}x \Bigg\{\delta\mu' + 3{\cal H}
\left(\delta{\cal P} + \delta\mu\right) -3 \left(\bar\rho + \bar p\right){\psi}' \nb\\
& & ~  
+ \frac{1}{2a^{4}}\Bigg[\Big(a^{3} \bar{J}_{A}\Big)'  \delta{A}
+ \hat{A}\Big(a^{3}\big(\delta{A} - 3 \hat{A}\psi\big)\Big)' \Bigg]\Bigg\} =
0,\nb\\
\\
\lb{4.9b}
 & & q'+3 {\cal H}q  - a\delta{\cal P} -
{2a\over3}\left( \vec\nabla ^{2}+3k \right)\Pi \nb\\
&& ~~~~~~~~~~~~~~~~~~~~~~~~~~~ 
+ \frac{1}{2}\bar{J}_{A} \delta{A} = 0, ~~
 \eqn
where $\bar{J}_{A}$  is given by Eq.(\ref{3.8b}).  

This completes the general description of linear scalar perturbations in the FRW background with any spatial 
curvature in the framework of the HMT setup with any given $\lambda$, generalized recently by da Silva \cite{Silva}.
  
\section{Stability of the Minkowski Spacetime}

\renewcommand{\theequation}{5.\arabic{equation}} \setcounter{equation}{0}

The stability of the maximal symmetric spacetimes in the da Silva generalization with $\lambda \not=1$ was considered in
\cite{Silva} with detailed balance condition. Since the Minkowski is not a solution of the theory when detailed
balance condition is imposed, so the analysis given in \cite{Silva} does not include the case where the Minkowski spacetime
is  the background. However, for the potential given by Eq.(\ref{2.5a}), the detailed balance condition is broken,
and the Minkowski spacetime now is a solution of the theory. Therefore, in this section we study the stability
of the Minkowski spacetime with any given $\lambda$. The case with $\lambda =  1$ was considered in Paper II,
so in this section we consider only the case with $\lambda \not= 1$. 

It   is easy to show  that the Minkowski spacetime, 
\bq
\lb{5.1}
a =1, \; \;\;
 \bar{A} = \bar{\varphi}   =  k = 0,  
 \eq
 is a solution of the da Silva generalization even with $\lambda \not=1$, provided that
 \bq
\lb{5.2}
   \Lambda_{g}=\Lambda    = \bar{J}_{A} = \bar{J}_{\varphi} = \bar{\rho} = \bar{p}
 = 0.
 \eq
 Then,   the linearized Hamiltonian constraint (\ref{4.4}) is satisfied identically, while the super-momentum constraint
  (\ref{4.5}) yields,
 \bq
 \lb{5.3}
\partial^{2}B = \frac{3\lambda - 1}{1-\lambda}\dot{\psi},
 \eq
where $\partial^{2} = \delta^{ij}\partial_{i}\partial_{j}$.  Eqs.(\ref{4.6a}) and (\ref{4.6b}) reduce to,
 \bqn
 \lb{5.4a}
& & \partial^{2}\Big(\partial^{2}B + 3 \dot{\psi}\Big) = 0,\\
 \lb{5.4b}
 & & \partial^2\psi = 0.
 \eqn
 Then, we have $\delta{F}_{ij}
= - \psi_{,ij}$, and the trace and traceless parts of the dynamical equations reduce, respectively, to
\bqn
\lb{5.5a}
\ddot{\psi} - \frac{2}{3(3\lambda -1)}\partial^{2}\delta{A} + \frac{1}{3}\partial^{2}\dot{B} = 0,\\
\lb{5.5b}
\dot{B} =  \delta{A} - \psi.
\eqn
It can be shown that Eqs.(\ref{5.4a}) and (\ref{5.5a}) are not independent, and can  be obtained from 
Eqs.(\ref{5.3}), (\ref{5.4b}) and (\ref{5.5b}). Eq.(\ref{5.4b}) shows  that $\psi$ is not propagating, and
with proper boundary conditions, we can set $\psi = 0$. Then, Eqs.(\ref{5.3}) and (\ref{5.5b}) show that $B$
and $\delta{A}$ are also not propagating, and shall also vanish with proper boundary
conditions. Therefore, we finally obtain
\bq
\lb{5.2a}
\psi = B =\delta{A} = 0. 
\eq 
Thus, the scalar perturbations   even with $\lambda \not=1$ vanish identically in the Minkowski 
background. Hence,  the spin-0 graviton is  eliminated
in the da Silva generalization  even  for any given coupling constant $\lambda$. 


\section{Ghost and Strong Coupling}

\renewcommand{\theequation}{6.\arabic{equation}} \setcounter{equation}{0}

To consider the ghost and strong coupling problems, we first note that they are closely related to the fact that
$\lambda \not=1$. The parts that depend on $\lambda$ are the kinetic part, ${\cal{L}}_{K}$,
 and the interaction part ${\cal{L}}_{\lambda}(K_{ij}, \varphi)$ between
the  extrinsic curvature  $K_{ij}$ and the Newtonian pre-potenital $\varphi$. With the gauge choice $\varphi = 0$, 
we can see that the latter vanishes identically. Then, it is sufficient to consider only the 
kinetic  part $S_{K}$, the IR terms $R$ and $\Lambda$,  and the source term $S_{M}$ \cite{KA,PS,WWb}, 
\bq
\lb{6.0}
S_{IR} = \int{dt d^{3}x N \sqrt{g}\Big({\cal{L}}_{K} + R - 2\Lambda + {\cal{L}}_{M}\Big)}.
\eq
Second,  the presence of  matter  is to produce non-zero perturbations. Otherwise, the spacetimes, to zero-order, are
the maximally symmetric spacetimes.  In these backgrounds,   when matter is not present, the corresponding
metric and gauge field perturbations, $\psi,\; B $ and $\delta{A}$, vanish identically,  as shown in the last section   for the Minkowski 
spacetime, and in \cite{Silva} for the (anti-) de Sitter one. 
On the other hand,  ${\cal{L}}_{M}$ does not depend on
$\lambda$, so it does not   contribute to the strong coupling and ghost problems. Therefore, the only role that 
${\cal{L}}_{M}$ plays here is to produce non-vanishing $\psi,\; B $ and $\delta{A}$. 
It is interesting to note that to study the strong coupling problem, in \cite{BPS}  the authors assumed that the background metric has non-vanishing 
extrinsic and spatial curvatures in the scale $L$:  $\bar{R}_{ij} \sim 1/L^{2}$ and $\bar{K}_{ij} \sim 1/L$, instead of non-vanishing $\psi$ and $B$ assumed 
here as well as in \cite{PS,KA}.
But, the purposes are the same: to provide a  environment so that the strong coupling problem can manifest itself properly, if it exists.
In  the following,  we shall consider two different kinds of gravitational fields: one represents spacetimes in which  the flat FRW universe with $\Lambda = 0$
can be considered as their  zero-order approximations;  and the other represents  static weak gravitational fields, in which the Minkowski spacetime can be considered 
as their zero-order approximations. 


\subsection{Ghost-free Conditions}

In the flat FRW background, the quadratic part of $S_{IR}$  is given by \cite{HWW}, 
\bqn
\lb{6.1}
  S^{(2)}_{IR} &=& \zeta^{2}\int{d\eta d^{3}x a^{2}\Bigg\{\big(1-3\lambda\big)\Big[3\psi'^{2} + 6{\cal{H}}\psi\psi' }\nb\\
& &  ~~~~~ + 2 \psi' \partial^{2}B  + \frac{9}{2}{\cal{H}}^{2}\psi^{2}\Big] + 2\big(\partial\psi\big)^{2} \nb\\
& & ~~~~~ + (1-\lambda) \big(\partial^{2}B\big)^{2}\Bigg\}.~~~
\eqn
Note that in writing the above expression, we had ignored the term, $ {\cal{L}}_{M}$,  as it has no contributions to both
the ghost and  the strong problems, as  mentioned above.   Then, from the super-momentum constraint (\ref{4.5}),  we find that
\bq
\lb{6.2}
\partial^2 B = \frac{3\lambda -1}{1-\lambda}\psi' - \frac{8\pi G a q}{1-\lambda}.
\eq
Substituting it into Eq.(\ref{6.1}), we obtain 
\bqn
\lb{6.3}
S^{(2)}_{IR} &=&  \zeta^{2} \int{d\eta d^3x a^{2(1+\delta)}\Bigg\{- \frac{2 }{c^{2}_{\psi}}{\tilde\psi}^{'2} + 2 \big(\partial{\tilde\psi}\big)^{2}}\nb\\
&& - \frac{9\lambda(3\lambda-1)}{2}{\cal{H}}^{2}{\tilde\psi}^2 + \frac{\tilde{q}^{2}}{c^{2}_{\psi}}\Bigg\},
\eqn
where 
\bq
\lb{6.3a}
c_{\psi}^{2} = \frac{1-\lambda}{3\lambda - 1},\;\;\;
\psi = a^{\delta}\tilde{\psi}, \;\;\;
q = \frac{\sqrt{3\lambda - 1} \tilde{q}}{8\pi G a^{1-\delta}},
\eq
and  $\delta \equiv - 3(1-\lambda)/2$. Thus, the ghost-free condition requires $c_{\psi}^{2} < 0$, or equivalently, 
\bq
\lb{6.4}
i) \; \lambda > 1, \;\;\; {\mbox{or}}\;\; ii) \;\;\; \lambda < \frac{1}{3},
\eq
which are precisely the conditions obtained in Paper I in the SVW setup \cite{WM}. 

It should be noted that the conditions (\ref{6.4}) also hold in the non-flat  FRW backgrounds,  as one can easily show by following the above arguments. 

In addition,  the expression of $S^{(2)}_{IR}$ given by Eq.(\ref{6.1}) is the same as that given in \cite{CHZ}, but  different 
from the   one given in \cite{CB}. After the typos of \cite{CB} are corrected, it can be shown that, in contrast
to their claims, the scalar modes are both ghost-free and stable in the ranges of $\lambda$ defined by Eq.(\ref{6.4}),   
when the matter field is a scalar and satisfies   the scalar field  equations.

\subsection{Strong Coupling Problem}


As mentioned previously, we shall consider two different kinds of spacetimes. In the following, let us consider them separately. 

\subsubsection{Flat FRW Background}

In this case we adopt the gauge,
\bq
\lb{6.5}
N = a,\;\;\; N_{i} = a^{2}e^{B} \partial_{i}{B},\;\;\; g_{ij} = a^{2}e^{-2\psi}\delta_{ij},
\eq
which reduces to the linear perturbations studied in the previous sections to the first order of $\psi$ and $B$.  This gauge is
slightly different from the one used in   \cite{WWb,KA,PS}. 
Then, we find
\bq
\lb{6.5a}
R = \frac{2 e^{2\psi}}{a^{2}}\Big(2 \partial^{2}\psi - \big(\partial\psi\big)^{2}\Big),
\eq
and 
the kinetic action $S_{K} $ is given by Eq.(\ref{A.1}). Hence, to the third-order of $\psi$ and $B$, we find that
\bqn
\lb{6.6}
S_{IR}^{(3)} &=& \zeta^{2} \int{d\eta d^{3}x a^{2}\Bigg\{2\psi\Big[\psi\partial^{2}\psi + \big(\partial\psi\big)^{2}\Big]}\nb\\
& & 
+ \frac{9}{2}\big(3\lambda - 1\big)\Big( 2 \psi{\psi'}^{2}  + 6{\cal{H}}\psi^{2}\psi'+ 3{\cal{H}}^{2}\psi^{3}\Big)\nb\\
& & + (3\lambda -1) \Big[2(\psi' + {\cal{H}}\psi)\big(\psi^{,k}B_{,k}\big)\nb\\
& & ~~~~~~~~~~~~~~~ + \psi(2\psi' + {\cal{H}}\psi)\partial^{2}B\Big]\nb\\
& & - 2\Big[(3\lambda -1){\cal{H}}B - (\lambda -1)\partial^{2}B\Big]\big(\psi^{,k}B_{,k}\big)\nb\\
& &   - 2\big(3\lambda -1\big)\big(\psi' + {\cal{H}}\psi\big)\Big[B\partial^{2}B + \big(\partial{B}\big)^{2}\Big]\nb\\
&& + 4\psi_{,k}B_{,l}B^{,kl}  + (\psi + 2B)\Big[B^{,kl}B_{,kl} - \lambda\big(\partial^{2}B\big)^{2}\Big]   \nb\\ 
%
%
%
%
& & + (3\lambda -1){\cal{H}} B\Big[B\partial^{2}B + 2\big(\partial{B}\big)^{2}\Big]\nb\\
&& - 2\lambda (\partial B\big)^{2}\partial^{2}B  + 2 B^{,kl}B_{,k}B_{,l}\Bigg\}.
\eqn

Following \cite{WWb}, we first write the quadratic action (\ref{6.3}) in its canonical form with order-one  coupling constants,  by
using the coordinate transformations,
\bq
\lb{6.7}
\eta = \alpha \hat{\eta},\;\;\; x^{i} = \alpha|c_{\psi}| \hat{x}^{i},
\eq
and redefinitions of  the canonical variables,
\bqn
\lb{6.8}
\tilde{\psi} &=& \frac{\hat{\psi}}{M_{pl}|c_{\psi}|^{1/2}\alpha},\nb\\
\tilde{q} &=& \frac{ \sqrt{2} \hat{q}}{  M_{pl}|c_{\psi}|^{1/2}\alpha^{2}}.
\eqn
It must not be confused with the constant $\alpha$ used here and the one used in the previous sections for the $U(1)$ gauge generator.
Then,   
from Eq.(\ref{6.2}) we find that
\bqn
\lb{6.10}
B &=& - \frac{1}{|c_{\psi}|^{2}\partial^{2}}\left(\psi' - \frac{8\pi G aq}{3\lambda -1}\right) = \frac{\hat{B}}{M_{pl} |c_{\psi}|^{1/2}}, \nb\\
\hat{B} &=&  - \frac{a^{\delta}}{\hat{\partial}^{2}}\left(\hat{\psi}^{*} + \delta\hat{\cal{H}}\hat{\psi} - \sqrt{\frac{2}{3\lambda - 1}}\hat{q}\right),  
\eqn
where $\hat{\psi}^{*} = \partial\hat{\psi}/\partial\hat{\eta},\; \hat{\cal{H}} = a^{*}/a$. 
Inserting Eqs.(\ref{6.7})-(\ref{6.10}) into Eq.(\ref{6.6}), we obtain
\bqn
\lb{6.12}
S_{IR}^{(3)} &=&  \frac{1}{2M_{pl}} \int{d\hat{\eta} d^{3}\hat{x} a^{2}\Bigg\{\frac{|c_{\psi}|^{3/2}}{\alpha}  \hat{\cal{L}}^{(3)}_{1}}\nb\\
&& +  \frac{1}{|c_{\psi}|^{1/2} \alpha}\Big(\hat{\cal{L}}^{(3)}_{2} + \hat{\cal{L}}^{(3)}_{3}  + \hat{\cal{L}}^{(3)}_{4}\Big) 
\nb\\
& & +   \frac{1}{|c_{\psi}|^{5/2}\alpha}\hat{\cal{L}}^{(3)}_{5}\nb\\
& & +  \frac{1}{|c_{\psi}|^{1/2}}\Big(\hat{\cal{L}}^{(3)}_{6} + \hat{\cal{L}}^{(3)}_{7}\Big)\nb\\
& &
+  \frac{1}{|c_{\psi}|^{5/2}}\Big(\hat{\cal{L}}^{(3)}_{8} + \hat{\cal{L}}^{(3)}_{9}\Big)\nb\\
& & +  \frac{\alpha}{|c_{\psi}|^{1/2}} \hat{\cal{L}}^{(3)}_{10}\Bigg\}, 
\eqn
where ${\cal{L}}^{(3)}_{i}$'s are  given in Eq.(\ref{A.4}). 
Clearly, for any chosen $\alpha$ some of the coefficients of  ${\cal{L}}^{(3)}_{i}$'s always become unbounded as $c_{\psi} \rightarrow 0$, 
that is, the corresponding theory is indeed plagued with the strong coupling problem. 

To study it further, let us consider the rescaling,
\bqn
\lb{6.14}
& & \hat{\eta} \rightarrow s^{-\gamma_{1}} \hat{\eta},\;\;\; 
\hat{x}^{i} \rightarrow s^{-\gamma_{2}} \hat{x}^{i},\nb\\
& &  \hat{\psi} \rightarrow s^{\gamma_{3}} \hat{\psi},\;\;\; 
\hat{q} \rightarrow s^{\gamma_{4}} \hat{q}.
\eqn
Then, $S^{(2)}_{IR}$ given by Eq.(\ref{6.3}) is invariant for $\gamma_{1} = \gamma_{2} = \gamma_{3} = \gamma_{4}/2  = \gamma$. Without loss of generality, one can set
$\gamma = 1$. For such a choice of $\gamma$, it can be shown that $\hat{B}$ is scale-invariant, 
\bq
\lb{6.15}
\hat{B} \rightarrow  \hat{B}.
\eq
Then,  in  $S^{(3)}_{IR}$ of Eq.(\ref{6.12})  the first five terms  are scaling as $s^{1}$, and the sixth to ninth terms  all scaling as $s^{0}$, while the last term is scaling
as $s^{-1}$.  Thus, all the first five terms are irrelevant in the low energy limit, and diverge  in the UV, so they are all not renormalizable \cite{Pol}. The sixth to ninth terms 
are marginal, and are strictly renormalizable, while the last term is relevant  and superrenormalizable. 
This indicates that the perturbations break 
down when the coupling coefficients greatly exceed units. To calculate these  coefficients, let us consider a process at the energy scale $E$. Then, we find that
 the ten terms in the cubic action $S^{(3)}_{IR}$ have, respectively, the  magnitudes, $(E, E, E, E, E, E^{0}, E^{0}, E^{0}, E^{0}, E^{-1})$,
  for example,
\bq
\lb{6.16}
\int{d\hat{\eta} d^{3}\hat{x}  \hat{\psi}^{*}  
\big(\hat{\partial}_{i}\hat\psi\big)\big(\hat{\partial}^{i}\hat{B}\big)} \simeq E.
\eq
Since the action is dimensionless, all the  coefficients in (\ref{6.12}) must have the dimensions $E^{-n_{s}}$, where $n_{s} =  (1, 1, 1, 1, 1, 0, 0, 0, 0, -1),\;
(s = 1, 2, 3, ..., 10)$.  Writing them in the form,
\bq
\lb{6.17}
\lambda_{s} = \left(\frac{\hat{\lambda}_{s}}{\Lambda_{s}}\right)^{n_{s}},
\eq
where $\hat{\lambda}_{s}$ is a dimensionless parameter of order one, one finds that $\Lambda_{s}$ for $s = 1, 2,  3, 4,  5, 10$ are given by Eq.(\ref{A.5a}). 
Translating it back to the coordinates $\eta$ and $x^{i}$, 
the energy and momentum scales are
given by
\bq
\lb{6.18}
\Lambda_{s}^{\omega} = \frac{\Lambda_{s}}{\alpha},\;\;\;
\Lambda_{s}^{k} = \frac{\Lambda_{s}}{\alpha|c_{\psi}|}.
\eq
For $s = 6, 7, 8, 9$, the coupling coefficients are given by   Eq.(\ref{A.5b}). From these expressions, one can see that
 the   lowest scale of $\Lambda_{s}^{\omega}$ and $\Lambda_{s}^{k}$'s is 
given by 
\bq
\lb{6.19}
\Lambda_{min} = \Lambda^{\omega}_{5} \simeq  |c_{\psi}|^{5/2}M_{pl},
\eq
as $c_{\psi} \rightarrow 0$. For any process with energy higher than it, the corresponding coupling constants become larger than unit, and then the   strong coupling problem
rises. 

Thus, to be consistent with observations in the IR,  $\lambda$ is required to be closed to its relativistic value $\lambda_{IR} = 1$. On the other hand,
to avoid the strong
coupling problem, the above shows that it cannot be too closed to it. 

\subsubsection{Static Weak Gravitational Fields}

When a static gravitational field produced by a source  is weak, such as the solar system, one can treat the problem as perturbations of the Minkowski spacetime.    
Since the Minkowski background is a particular case of the flat FRW spacetime,  one can consider its perturbations
still  given by Eq.(\ref{6.5}) but now with $a = 1$.  Due to the presence of matter, $\psi$ now is in general different from zero. Then, we find that
\bq
\lb{6.3aa}
S^{(2)}_{IR} =  \zeta^{2} \int{dt d^3x\Bigg(2 \big(\partial{\psi}\big)^{2}} - \frac{(8\pi Gq)^{2}}{\lambda - 1}\Bigg),
\eq
where
\bq
\lb{6.6b}
B = \frac{8\pi G}{(\lambda -1)\partial^{2}}q.
\eq
Setting 
\bqn
\lb{6.6c}
t &=& \alpha \hat{t},\;\;\; x^{i} = \alpha\hat{x}^{i},\nb\\
\psi &=& \frac{\hat{\psi}}{\sqrt{2} \zeta \alpha},\;\;\;
q = \frac{\sqrt{3\lambda -1}|c_{\psi}| \hat{q}}{8\pi G\zeta   \alpha^{2}},
\eqn
we find that $S^{(2)}$ given by Eq.(\ref{6.3aa}) takes its canonical form,
\bq
\lb{6.3ab}
S^{(2)}_{IR} =    \int{d\hat{t} d^3\hat{x}\Big(\big(\hat{\partial}{\hat\psi}\big)^{2} - \hat{q}^{2}\Big)}.
\eq

On the other hand, we have
\bqn
\lb{6.6a}
S_{IR}^{(3)} &=& \zeta^{2} \int{dt d^{3}x \Bigg\{2\psi\Big[\psi\partial^{2}\psi + \big(\partial\psi\big)^{2}\Big]}\nb\\
& & + 2 (\lambda -1)\partial^{2}B\Big(\psi^{,k}B_{,k}\Big) + 4\psi_{,k}B_{,l}B^{,kl}\nb\\
&&   + (\psi + 2B)\Big[B^{,kl}B_{,kl} - \lambda\big(\partial^{2}B\big)^{2}\Big]   \nb\\ 
&& - 2\lambda (\partial B\big)^{2}\partial^{2}B  + 2 B^{,kl}B_{,k}B_{,l}\Bigg\}\nb\\
&=& \frac{1}{M_{pl}}\int{d\hat{t}d^{3}\hat{x}\Bigg\{\frac{L^{(3)}_{1}}{\alpha} + \frac{2L^{(3)}_{2}}{\alpha}}\nb\\
& & + \frac{L^{(3)}_{3}}{(3\lambda -1)|c_{\psi}|^{2}\alpha}\nb\\ 
& &  
+ \left(\frac{2}{3\lambda -1}\right)^{3/2}   \frac{L^{(3)}_{4}}{|c_{\psi}|^{3}}\Bigg\},
\eqn
where
\bqn
\lb{6.6ac}
L^{(3)}_{1} &=& \hat{\psi}\Big[\hat{\psi}\big(\hat{\partial}^{2}\hat{\psi}\big) + \big(\hat{\partial}\hat{\psi}\big)^{2}\Big],\nb\\
L^{(3)}_{2} &=&  \big(\hat{\partial}^{2}\hat{B}\big) \big(\hat{\partial}^{k}\hat{\psi}\big)\big(\hat{\partial}_{k}\hat{B}\big),\nb\\
L^{(3)}_{3} &=&  \hat{\psi}\Big[\big(\hat{\partial}_{k}\hat{\partial}_{l}\hat{B}\big)^{2} - \lambda \big(\hat{\partial}^{2}\hat{B}\big)^{2}\Big]\nb\\
& &  +  4\big(\hat{\partial}^{k}\hat{\psi}\big)\big(\hat{\partial}^{l}\hat{B}\big)\big(\hat{\partial}_{k}\hat{\partial}_{l}\hat{B}\big) ,\nb\\
L^{(3)}_{4} &=&  \hat{B}\Big[\big(\hat{\partial}_{k}\hat{\partial}_{l}\hat{B}\big)^{2} - \lambda \big(\hat{\partial}^{2}\hat{B}\big)^{2}\Big]
   - \lambda \big(\hat{\partial}^{2}\hat{B}\big)  \big(\hat{\partial}\hat{B}\big)^{2}\nb\\
& &  +  \big(\hat{\partial}^{k}\hat{B}\big)\big(\hat{\partial}^{l}\hat{B}\big)\big(\hat{\partial}_{k}\hat{\partial}_{l}\hat{B}\big),
\eqn
but now with 
\bq
\lb{6.6ad}
B = \frac{{1}}{\zeta |c_{\psi}|\sqrt{3\lambda -1}}\left(\frac{1}{\hat{\partial}^{2}}\hat{q}\right) \equiv  \frac{\hat{B}}{\zeta|c_{\psi}|\sqrt{3\lambda -1}}.
\eq

Considering the rescaling (\ref{6.14}) with $\hat{t} = \hat{\eta}$, we find that $S^{(2)}_{IR}$ given by Eq.(\ref{6.3ab}) is invariant, provided that
$\gamma_{3} = (\gamma_{1} + \gamma_{2})/2$ and $\gamma_{4} = (\gamma_{1} + 3\gamma_{2})/2$. Without loss of generality, we can set
$\gamma_1 = \gamma_2 = 1$, and then $\hat{B}$ scales exactly as that given by Eq.(\ref{6.15}), while the four terms in $S^{(3)}_{IR}$ of 
Eq.(\ref{6.6a}) scale, respectively, as $s^{1},\; s^{1},\; s^{1}$ and $s^{0}$. Then, following the analysis given between Eqs.(\ref{6.17}) and
(\ref{6.19}), we find that $\Lambda^{k}_{s} = \Lambda^{\omega}_{s}$ for $s = 1, 2, 3$, where
\bqn
\lb{6.6ae}
\Lambda^{\omega}_{1} &=&   2 \Lambda^{\omega}_{2} = M_{pl},\nb\\
\Lambda^{\omega}_{3} &=& (3\lambda -1)M_{pl}|c_{\psi}|^{2},
\eqn
and
\bq
\lb{6.6af}
\lambda_{4} =  \left(\frac{2}{3\lambda -1}\right)^{3/2}\frac{1}{M_{pl}|c_{\psi}|^{3}}.
\eq
 Clearly, as $\lambda \rightarrow 1$, the coupling  also becomes strong.  In particular, since the fourth term scales as $s^{0}$,  its amplitude   remain the same,
 as the energy scale of the system changes.  That is, this term is equally important at all energy scales.  The strength of this term gives  the lowest  energy scale,
  as  $c_{\psi} \rightarrow 0$. Therefore, now  we have
 \bq
 \lb{6.6ag}
 \Lambda_{min} \simeq M_{pl}|c_{\psi}|^{3}.
 \eq
 
It should be noted that, in the above   we studied the strong coupling problem   only in terms of $\psi$. 
Then, one may argue that our above conclusions may be gauge-dependent. In the following,  we shall show that this is not true. 

Let us first note that in the static case the gauge invariant quantity $\Psi$ is precisely equal to  $\psi$, as one can see from Eq.(\ref{4.0f}). Therefore, in this case
the coupling  indeed becomes strong when $ E > M_{pl}|c_{\psi}|^{3}$, even in terms of the gauge-invariant quantity.  

On the other hand,  in the cosmological case, 
from Eqs.(\ref{4.0f}) and (\ref{6.10})  we find that the gauge-invariant quantity $\Psi$   can be written as
\bq
\lb{6.6ah}
\Psi = \frac{a^{\delta}\hat{\Psi}}{M_{pl}|c_{\psi}|^{1/2}\alpha},
\eq
where
\bq
\lb{6.6ai}
\hat{\Psi} \equiv \hat{\psi} + \frac{\alpha{\cal{H}}}{\hat{\partial}^{2}}\Bigg[\alpha\big(\hat{\psi}' + \delta{\cal{H}}\hat{\psi}\big)
- \sqrt{\frac{2}{3\lambda -1}} \hat{q}\Bigg].
\eq
Since the lowest energy scale (\ref{6.19}) is independent of $\alpha$ (as it should be), we can always choose 
$\alpha \propto |c_{\psi}|^{d}, (d > 0)$, so that $\hat{\Psi} \simeq \hat{\psi}$ and ${\Psi} \simeq {\psi}$ as $|c_{\psi}| \rightarrow 0$.  Then, one can repeat the analysis
in terms of ${\Psi}$ and $\hat{\Psi}$ and finds that the strong coupling problem exists even in terms of $\Psi$, which is gauge-invariant. 

\section{Conclusions}

\renewcommand{\theequation}{7.\arabic{equation}} \setcounter{equation}{0}

Recently, Horava and Melby-Thompson \cite{HMT} proposed a new  version of the HL theory of gravity, 
in which the spin-0 graviton, appearing in all the previous versions of the HL theory,  is eliminated by introducing 
a Newtonian pre-potential $\varphi$ and a local $U(1)$ gauge field $A$. Such a setup was orginally believed valid only
for $\lambda =1$. However, da Silva  argued that the HMT setup can be
easily generalized to the case with $\lambda \not=1$. With such a generalization, the three challenging questions,
ghost, stability and strong coupling, all related with $\lambda \not=1$ and plagued  in most of the previous versions 
of the HL theory \cite{Mukc,Sotiriou},  rise again.

In this paper,  we  addressed these issues, by first developing the linear scalar perturbations of the FRW spacetimes
for any given $\lambda$ in the da Silva generalization. In particular, in Sec. II we  derived all the field equations and the corresponding conservation laws,
while in Sec. III we studied the cosmological models of the FRW universe with any given spatial curvature $k$. 
When $\bar{J}_{A} = 0$, from Eq.(\ref{3.8b}) we find that $k = 0 = \Lambda_{g}$,  that is,  the universe must be flat.  
When the matter is described by a scalar  field, one can see that  $\bar{J}_{A}$ indeed vanishes. Therefore, in all
the inflationary models described by a scalar field, the FRW universe is necessarily  flat.  Thus,  the theory naturally 
gives rise to a flat FRW universe, which is consistent with all the observations carried out so far \cite{WMAP}.  

Then, in Sec. IV we presented the general formulas for the linear scalar perturbations. By studying the general gauge transformations
of $U(1) \ltimes {\mbox{Diff}}(M,{\cal{F}})$, we found that there are only three gauge-invaraint quantities, and constructed them
explicitly, as given by Eq.(\ref{4.0f}). Applying these formulas to the Minkowski background,  in Sec. V we showed explicitly that the 
Minkowski spacetime is stable, and the corresponding spin-0 graviton is eliminated by the gauge field even for $\lambda \not= 1$.

In Sec. VI, we considered the ghost and strong coupling problems. To study them, we need to consider the cases where the linear perturbations
of the metric, described by $\psi$ and $B$ in the quasilongitudinal   gauge (\ref{4.0g}), are different from zero, so that these problems can manifest 
themselves, if they exist. 
One way to have non-vanishing  $\psi$ and $B$
is to assume that the spacetimes are not vacuum. In particular, taking the flat FRW universe as the background, we   found that the ghost-free conditions 
are the same as these found in Paper I in the SVW setup, given explicitly by Eq.(\ref{6.4}).  In such backgrounds, we found that
the strong coupling problem also shows up. In particular, for a process with energy $E$ higher than $ |c_{\psi}|^{5/2}M_{pl}$,  the corresponding 
coupling constants become much larger than unit, and then the   strong coupling problem rises.  In the static case, strong coupling problem also exists
for $E > |c_{\psi}|^{3}M_{pl}$.  
To resolve this problem, one way is to provoke the   Vainshtein
mechanism \cite{Vain}, similar to what was done previously in spherical static spacetimes \cite{Mukc}, as well as in cosmology \cite{WWb}, or use the 
BPS mechanism \cite{BPS,WWb}. 

The gauge field $A$ and the Newtonian pre-potentail $\varphi$ have no contributions to the vector and tensor perturbations, so
the results presented in \cite{Wang} in the SVW setup can be equally applied to the  da Silva generalization even with $\lambda \not=1$. In particular,  
it was shown that the vector perturbations vanish identically in the Minkowski background. 
Combining it with the result obtained in this paper, one can see that
 the only non-vanishing  part is the tensor one. 
As a result,  in the Minkowski background the gravitational sector is still described  
 only by the spin-2 massless graviton   even in  the da Silva generalization ($\lambda \not= 1$). 

Finally, we would like to note that, although this new version of the HL theory has several attractive features and solves various important 
issues plagued  in the previous versions,    many fundamental issues  still need  to be addressed, 
before it is considered as  a viable theory. These  include the strong coupling problem found above, 
the RG flow, phenomenological constraints  from the solar system tests,  the couplings of  matter fields to gravity, and so on.

~\\{\bf Acknowledgements:}  The authors thank   Robert  Brandenberger,  Shinji Mukohyama, Tony Padilla and Thomas Sotiriou for valuable comments and suggestions. 
The work of AW was 
supported in part by DOE  Grant, DE-FG02-10ER41692.

 \section*{Appendix: The kinetic action $S_{K}$}

 \renewcommand{\theequation}{A.\arabic{equation}} \setcounter{equation}{0}

 For the metric given by Eq.(\ref{6.5}), 
it can be shown that the kinetic part of the action is given by
\bqn
\lb{A.1}
S_{K} &=& \zeta^{2} \int{d\eta d^{3}x a^{2}\Bigg\{3\big(1 - 3\lambda\big)\big({\cal{H}} - \psi'\big)^{2}  e^{-3\psi}}\nb\\
& & + \Big[\big(B^{,ij} + B^{,i}B^{,j}\big)\big(B_{,ij} + B_{,i}B_{,j}\big)\nb\\
& & ~~~~ - \lambda\Big(\partial^{2}B + \big(\partial{B}\big)^{2}\Big)\Big]e^{2B + \psi}  \nb\\
& & - \lambda \big(\psi^{,k}B_{,k}\big)^{2} e^{2B+\psi}\nb\\
& & - 2\big(1-\lambda\big) \Big[\partial^{2}B + \big(\partial{B}\big)^{2}\Big] \big(\psi^{,k}B_{,k}\big) e^{2B+\psi}\nb\\
& & - 2\big(1-3\lambda\big)\big({\cal{H}} - \psi'\big)\big[\partial^{2}B + \big(\partial{B}\big)^{2} \nb\\
& & ~~~~~~~~~~~~~~~~~~~~~~~~~~~~ - \big(\psi^{,k}B_{,k}\big)\Big] e^{B - \psi} \nb\\
& & + 4\psi_{,i}B_{,j}\big(B^{,ij} + B^{,i}B^{,j}\big) e^{2B+\psi}\nb\\
& & + \Big[2\big(\partial\psi\big)^{2} \big(\partial{B}\big)^{2} + \big(\psi^{,k}B_{,k}\big)^{2}\Big]e^{2B+\psi}\Bigg\},
\eqn
from which we find that   the quadratic part is given by
\bqn
\lb{A.2}
  S^{(2)}_{K} &=& \zeta^{2}\int{d\eta d^{3}x a^{2}\Big\{\big(1-3\lambda\big)\Big[3\psi'^{2} + 18{\cal{H}}\psi\psi' }\nb\\
& &  + 2 \psi' \partial^{2}B  + \frac{27}{2}{\cal{H}}^{2}\psi^{2}\Big] + (1-\lambda) B\partial^{4}B\Big\}.~~~
\eqn
This is different from the expression given by Eq.(\ref{6.1}). The reason is that, in the calculations of Eq.(\ref{6.1}), the
3-metric $g_{ij}$ is approximated to the first-orders of $\psi$ and $B$, as one can see from Eq.(\ref{4.0a}), while
 $g^{ij}$ to their second orders (So does $\sqrt{g}$). For detail, we refer readers to \cite{RT}. However, in the derivation
 of Eq.(\ref{A.2}), we practically expanded both  $g_{ij}$ and  $g^{ij}$ to second orders. It is interesting to note that
 this difference does not affect the super-momentum constraint (\ref{6.2}), which can be also obtained by the variation
 of $S^{(2)}_{IR}$ with respect to $B$. Since the $B$-terms in both expressions of Eqs.(\ref{6.1}) and (\ref{A.2}) are the same,
 so is the resulting equation obtained by the   variation of $S^{(2)}_{IR}$ with respect to $B$.
 
 Substituting Eq.(\ref{6.2}) into Eq.(\ref{A.2}), we find that 
\bqn
\lb{A.3}
S^{(2)}_{IR} &=&  \zeta^{2} \int{d\eta d^3x a^{2(1+\delta)}\Bigg\{- \frac{2 }{c^{2}_{\psi}}{\tilde\psi}^{'2} + 2 \big(\partial{\tilde\psi}\big)^{2}}\nb\\
&& - \frac{27(3\lambda -2)(3\lambda-1)}{2}{\cal{H}}^{2}{\tilde\psi}^2 + \frac{\tilde{q}^{2}}{c^{2}_{\psi}}\Bigg\},
\eqn
where $S^{(2)}_{IR} = S^{(2)}_{K} + S^{(2)}_{R}$, and $c_{\psi}, \; \tilde{\psi}$ and $\tilde{q}$ are defined by Eq.(\ref{6.3a}) but now with $\delta = 9(\lambda -1)/2$. Then,
we find that the ghost-free conditions are the same as that given by Eq.(\ref{6.4}). 

One can show that the conclusions regarding to the strong coupling problem are also independent of the use of either the expression
(\ref{A.3}) or (\ref{6.3}) for $S^{(2)}_{IR}$. 

Inserting Eqs.(\ref{6.7})-(\ref{6.10}) into Eq.(\ref{6.6}), we find that $S^{(3)}_{IR}$ is given by Eq.(\ref{6.12}), where
\bqn
\lb{A.4}
 \hat{\cal{L}}^{(3)}_{1} &=& \frac{9}{2}(3\lambda -1)a^{3\delta}\Big[2\hat{\psi}\big(\hat{\psi}^{*} + \delta \hat{\cal{H}}\hat{\psi}\big)^{2}\nb\\
 & & ~~~~
 + 6 \hat{\cal{H}}\hat{\psi}^{2}\big(\hat{\psi}^{*} + \delta \hat{\cal{H}}\hat{\psi}\big) + 3  \hat{\cal{H}}^{2}\hat{\psi}^{3}\Big],\nb\\ 
\hat{\cal{L}}^{(3)}_{2} &=&2a^{3\delta}\Big[\hat{\psi}^{2} \hat{\partial}^{2}\hat{\psi} + \hat{\psi}\big(\hat\partial\hat{\psi}\big)^{2}\Big],\nb\\
\hat{\cal{L}}^{(3)}_{3} &=&\big(3\lambda - 1\big)a^{2\delta}\Bigg\{2\Big[\hat{\psi}^{*} + (1 + \delta) \hat{\cal{H}}\hat{\psi}\Big]
\big(\hat{\partial}_{i}\hat\psi\big)\big(\hat{\partial}^{i}\hat{B}\big)\nb\\
   & & ~~~~~~~~~ +\hat{\psi} \Big[2\hat{\psi}^{*} + (1 + 2\delta) \hat{\cal{H}}\hat{\psi}\Big] \big(\hat{\partial}^{2}\hat{B}\big)\Bigg\},\nb\\
\hat{\cal{L}}^{(3)}_{4} &=&2(3\lambda -1)a^{\delta}    \big(\hat{\partial}^{2}\hat{B}\big)  \big(\hat{\partial}_{i}\hat\psi\big)\big(\hat{\partial}^{i}\hat{B}\big),\nb\\
 \hat{\cal{L}}^{(3)}_{5} &=& a^{\delta} \Big[4\big(\hat{\partial}^{k}\hat{\partial}^{l}\hat{B}\big) \big(\hat{\partial}_{k}\hat{\psi}\big) \big(\hat{\partial}_{l}\hat{B}\big)  
+ \big(\hat{\partial}^{k}\hat{\partial}^{l}\hat{B}\big)^{2} \nb\\
& & ~~~~~ - \lambda  \big(\hat{\partial}^{2}\hat{B}\big)^{2}\Big],\nb\\
\hat{\cal{L}}^{(3)}_{6} &=&2(1 - 3\lambda)a^{\delta}  \Big[\hat{\psi}^{*} + (1 + \delta) \hat{\cal{H}}\hat{\psi}\Big]\times\nb\\
& & ~~~~~~~~
         \Big[\hat{B}\big(\hat{\partial}^{2}\hat{B}\big)  +   \big(\hat{\partial}\hat{B}\big)^{2}\Big],\nb\\    
\hat{\cal{L}}^{(3)}_{7} &=&2(1 - 3\lambda)a^{\delta} \hat{\cal{H}}\hat{B}\big(\hat{\partial}_{i}\hat\psi\big)\big(\hat{\partial}^{i}\hat{B}\big),\nb\\         
\hat{\cal{L}}^{(3)}_{8} &=& 2\hat{B} \Big[\big(\hat{\partial}^{k}\hat{\partial}^{l}\hat{B}\big)^{2}   - \lambda  \big(\hat{\partial}^{2}\hat{B}\big)^{2}\Big],\nb\\
\hat{\cal{L}}^{(3)}_{9} &=& 2 \Big[\big(\hat{\partial}^{k}\hat{\partial}^{l}\hat{B}\big) \big(\hat{\partial}_{k}\hat{B}\big) \big(\hat{\partial}_{l}\hat{B}\big)  
 - \lambda  \big(\hat{\partial}\hat{B}\big)^{2} \big(\hat{\partial}^{2}\hat{B}\big)\Big],\nb\\
 \hat{\cal{L}}^{(3)}_{10} &=&(3\lambda -1) \hat{\cal{H}}\hat{B} \Big[\hat{B}\big(\hat{\partial}^{2}\hat{B}\big)  +   2\big(\hat{\partial}\hat{B}\big)^{2}\Big].
 \eqn

The coupling coefficients of these terms defined by Eq.(\ref{6.17}) for $s = 1, 2, 3, 4, 5, 10$ are given by
\bqn
\lb{A.5a}
\Lambda_{1}  &=& \frac{4M_{pl}\alpha}{9(3\lambda -1) |c_{\psi}|^{3/2}}, \;\;\;
\Lambda_{2}  = M_{pl}  |c_{\psi}|^{1/2}\alpha, \nb\\
\Lambda_{3} &=& \frac{2M_{pl} |c_{\psi}|^{1/2}\alpha}{3\lambda -1},\;\;\;
\Lambda_{4} = \frac{M_{pl} |c_{\psi}|^{1/2}\alpha}{3\lambda -1},\nb\\
\Lambda_{5}  &=& 2M_{pl}  |c_{\psi}|^{5/2}\alpha, \;\;\;
\Lambda_{10} = \frac{(3\lambda -1)\alpha}{2 |c_{\psi}|^{1/2}M_{pl}}.
\eqn
We also have
\bqn
\lb{A.5b}
\lambda_{6} &=& \lambda_{7} =  \frac{3\lambda -1}{|c_{\psi}|^{1/2}M_{pl}},\nb\\
\lambda_{8} &=& \lambda_{9} =  \frac{ 1}{|c_{\psi}|^{5/2}M_{pl}}.
\eqn


\end{document}